\documentclass{ws-procs9x6}                                    
\usepackage{epsfig}       
\usepackage{amssymb}     
\usepackage{amsfonts}        
            
\newcommand{\be}{\begin{equation}}
\newcommand{\ee}{\end{equation}}
\newcommand{\bea}{\begin{eqnarray}}  
\newcommand{\eea}{\end{eqnarray}}
\newcommand{\beas}{\begin{eqnarray*}}
\newcommand{\eeas}{\end{eqnarray*}}
  
\newcommand{\non}{\nonumber}
\newcommand{\bquo}{\begin{quote}}
\newcommand{\enqu}{\end{quote}}

\def\de{\partial}
\def\Tr{ \hbox{\rm Tr}}

\def\o{\over}

\def\bra{\langle}
\def\ket{\rangle}

\def\2{{1\over 2}}

\begin{document}

\title{Quantum Nonabelian Monopoles}

\author{K. Konishi} 
  
\address {Dipartimento di Fisica,``E. Fermi"\\
Universit\`a di Pisa, \\ 
Via Buonarroti, 2, Ed. C\\
56127 Pisa, Italy \\ 
E-mail: konishi@df.unipi.it}


\maketitle   

\abstracts{   We discuss  quantum mechanical and topological aspects of nonabelian monopoles.  Related recent results  on nonabelian vortices are also mentioned. }

 \section {Prologue}
 
 There are several reasons  to be interested in {\it quantum,  nonabelian}  monopoles. 
 First, if confinement of QCD is a sort  of dual superconductor,  it is  likely  to be one of nonabelian variety.
 Then the effective degrees of freedom involve nonabelian, and not,  better understood abelian monopoles.   Second, the phenomenon of confinement has  to do with fully quantum mechanical, and not semi-classical,  behavior of  the monopoles. Thirdly, the very concept  of  nonabelian monopoles   is, as we shall see,  intrinsically quantum mechanical, in contrast to that  of the 't Hooft-Polyakov monopole carrying an abelian charge only.  
A semi-classical consideration only might  easily lead us astray.   Finally, some recent 
 developments on nonabelian BPS {\it vortices} provide further hints on the subtle nature of 
 nonabelian monopoles and related dual gauge transformations. 
These considerations are sufficient motivations  to give a renewed look on the topological as well as dynamical aspects of these soliton states, in particular in relation to $N=2$  gauge theories. 

 \section { Confinement in    $SU(N)$   YM   Theory \label{sec:dualsc} } 
  
 The test charges in $SU(N)$   YM theory take values in   ($Z_N^{(M)}, Z_N^{(E)}$)  where $Z_N$  is the center of $SU(N)$ and  
$Z_N^{(M)}, Z_N^{(E)}$ refer to the magnetic and electric center charges. 
  ($Z_N^{(M)}, Z_N^{(E)}$)  classification of phases  follows \cite{tHooft,TH}. Namely, 

\begin{enumerate}    
 \item    If a field with  $x=(a, b)$ condenses,   particles $X= (A,B)$ with  
$$    \langle x, X \rangle   \equiv     a \, B - b \, A \ne 0 \quad  (mod \,\, N)   $$
are confined.    (e.g.   $\langle \phi_{(0,1)}\rangle \ne 0$ $\to$   Higgs phase.)
\item  Quarks are confined if some magnetically charged  particle $ \chi$  condenses, 
   $\langle \chi_{(1,0)}\rangle \ne 0.$  
\item  In the   softly broken $N=4$  (to $N=1$) theory   (often referred to as $N=1^*$)    all different types of massive vacua,
  related by $SL(2,Z)$, appear;  the  chiral condensates in each vacua are  known. 
\item 
{\it   Confinement index} \cite{CSW}   is   equal to  the smallest possible  $r \in Z_N^{(E)} $   for which  Wilson loop displays   no area law. 
For instance,  for   $SU(N)$  YM,      $r=N$   in the vacuum with complete confinement;     $r=1$  in the 
totally Higgs vacuum, etc.  

\item   In softly broken $N=2$  gauge theories,  dynamics turns out to be  particularly transparent.  

\end{enumerate}    

We are particularly interested in questions such as: 
 What is   $\chi $ in QCD?    How do they interact?   Is  chiral symmetry breaking  related to confinement?

 A familiar idea is that the ground state of QCD is a dual superconductor \cite{TH}.  
 Although there exist  no  elementary nor soliton  monopoles in QCD,   
monopoles  can be detected  as topological singularities (lines in $4D$)  of Abelian gauge fixing,  $SU(3) \to U(1)^2$, as suggested by 't Hooft.    Alternatively, one can assume that  certain  configurations  close  to the Wu-Yang monopole  
 ($SU(2)$)
  $$A_{\mu}^a \sim    (
\partial_{\mu} { n}\times
  { n})^a  + \ldots, \qquad      { n}({ r})=   { {  r} \over r}  \quad  
\Rightarrow  \quad       A_i^a  = \epsilon_{aij}{r^j \over r^3}  $$
dominate \cite{FN}.
  
 Although there is  some evidence  in lattice QCD   \cite{DG}
for ``Abelian dominance",  there remain several questions to be answered.   
 Do abelian  monopoles carry flavor?  What is   ${ L}_{eff}$?
What about  the  gauge dependence of such abelian gauge-fixed action? 
Most significantly,  does  dynamical  $SU(N) \to U(1)^{N-1}$ breaking occur?  That would imply  a
 richer spectrum of mesons  ($T_1 \ne T_2, $ etc)  not seen in Nature and not expected in QCD.  Both in Nature and presumably in QCD   there is  only one ``meson" state,
$   \sum_{i=1}^N  |\, q_i \,  {\bar q}_i \, \rangle,   $ 
{\it i.e.},  $1$  state  vs   $\left[{N \over 2}\right]$  states.  
Note that it is not sufficient to assume the symmetry breaking   $SU(N) \to U(1)^{N-1}\times  Weyl$ symmetry,
  with an extra discrete symmetry,   to solve the problem: the multiplicity would be wrong.   If nonabelian degrees of freedom are important,  after all, how do they manifest themselves?

 \section {``Semiclassical''  Nonabelian Monopoles} 

Let us review briefly the standard  results about  nonabelian monopoles \cite{Lb}-\cite{CJH}.  One is interested in a  system with gauge symmetry breaking 
\[ G   \,\,\,{\stackrel {\bra \phi \ket    \ne 0} {\longrightarrow}}     \,\,\, H   \]
 where $H$ is non abelian.  Asymptotic behavior of  scalar and gauge fields (for a finite action) are: 
 \[  
\phi \sim   U \cdot  \bra \phi \ket  \cdot U^{-1}  \sim       \Pi_2(G/H)  =
\Pi_1(H);     \label{eq1}
\]
\[   A_i^a  \sim  U \cdot {\de_i  }  U^{\dagger}   \to       F_{ij}   \sim      \epsilon_{ijk}  { r_k 
\o     r^3}  ({ \beta} \cdot  {\bf H}),
\qquad     H_i  \in  {\hbox {\rm  Cartan S.A.  of }} \, \,\,  G. 
\]
 Topological Quantization then leads to  
\be 2 \, {\mathbf \alpha }  \cdot {\mathbf \beta  } \subset {\mathbb Z}, \qquad  cfr. \quad  2 \, g_e    \,  g_m   = { n }  
  \ee
\[  \beta_i  =   {\hbox {\rm  weight vectors  of}} \,\,  {\tilde H} ~~~~(=  {\hbox {\rm  dual   of}} \,\, H),   
\]
namely, the nonabelian monopoles are characterized by the weight vectors of the dual group $ {\tilde H}.$
A general formula for the semiclassical monopole solutions   (set $\bra \phi_0 \ket   =  {\bf h} \cdot  {\bf H} $)
is   given in terms of  various broken $SU(2)$ subgroups,  
 { \footnotesize 
 \[      S_1=  { 1\o \sqrt{ 2 { \bf \alpha}^2}  }  (  E_{{ \bf \alpha}}   +   E_{-{ \bf \alpha}}     ); \qquad  
 S_2=  - { i \o \sqrt{ 2 {\bf \alpha}^2  }    }(  E_{{ \bf \alpha}}   -     E_{-{ \bf \alpha}} ); \qquad  
S_3=   {\bf \alpha}^{*} \cdot  {\bf H }; 
\label{su2g}\]} the nonabelian monopoles are basically  an  embedding of  the 't Hooft-Polyakov monopoles \cite{TP}  in such $SU(2)$ subgroups:
 \be   A_i({\bf r})  =  A_i^a({\bf r},  {\bf h} \cdot {\bf \alpha}) \, S_a;  \quad \phi({\bf r}) =   
  \chi^a({\bf r},  {\bf h} \cdot {\bf
\alpha})
\, S_a   +  [ \,  {\bf h}   -   ({\bf h}
\cdot {\bf \alpha}) \,    {\bf \alpha}^{*}  ]  
\cdot {\bf H},
\label{NAmonop}\ee
 where  ($\alpha^*\equiv  \alpha/(\alpha \cdot \alpha)$)
\[   A_i^a({\bf r}) =  \epsilon_{aij}  { r^j \o r^2}  A(r); \qquad   \chi^a({\bf r}) =  { r^a \o r} \chi(r), \qquad    \chi(\infty)=
  {\bf h} \cdot {\bf \alpha}.
\]
 The mass and $U(1)$ flux can be   easily  calculated:
 \[    M=\int d{\bf S} \, \cdot    \Tr  \, \phi \,  {\bf B}, \qquad   {\bf B}=  {    r_i  ({\bf S}\cdot {\bf r})  \o r^4}=  {  {\bf r}    S_3    \o r^3}  = {  {\bf
r}      \o r^3} \,  {\bf \alpha}^{*} \cdot  {\bf H};   \label{mfield} \]
$U(1)$  flux (for instance, for  $SU(N+1) \to  SU(N) \times U(1)$) is 
\be   F_m=  \int_{S^2}   d{\bf S}    \cdot   {  \Tr  \,  \phi \,  {\bf  B}  \o { 1\o \sqrt {2} }  ( \Tr \phi^2)^{1/2} }   \equiv 4 \pi \, g_m  =   2 \pi   \cdot
\sqrt{2 (N+1) \o N.  }
\label{monoflux} 
\ee
Example of dual groups  (defined by    $\alpha \Leftrightarrow  \alpha^*$) are:
\smallskip
\begin{table}[h]   
\begin{center}
\begin{tabular}{c  c   c}
\hline  
$SU(N)/Z_N       $        &   $\Leftrightarrow$                 &    $SU(N)     $          \\
 $ SO(2N)  $     &   $\Leftrightarrow$    &   $SO(2N) $       \\  
 $ SO(2N+1)  $     &   $\Leftrightarrow$     &   $USp(2N) $       \\ \hline
\end{tabular} 
 \vskip .3cm
\end{center}
\end{table}  
 
 \section{Some Examples }
 
 The simplest system with nonabelian monopoles involves the gauge symmetry breaking, 
 $$     SU(3) {\stackrel {\bra \phi \ket } {\longrightarrow}}     { SU(2) \times  U(1) \o {\mathbb Z}_2 }  , \qquad   \bra \phi\ket = 
 \pmatrix{  v & 0& 0  \cr  0 & v & 0 \cr  0&0& -2v  } 
$$  
 The monopole solutions are 
\bea   \phi ({\bf r})  &=& 
   \left( \begin{array}{ccc}
     -\frac{1}{2}v&0&0\\
     0&v&0\\
     0&0&-\frac{1}{2}v\\
   \end{array} \right)  +  3\,  v {\hat S} 
   \cdot \hat{r} \phi(r),  \qquad  
  \vec{A}({\bf r}) =    {\hat S} 
   \wedge \hat{r} A(r), \non  \label{su3sol}   \eea
 $\phi(r)$ and   $A(r)$ are BPS  't Hooft's  functions   with  $  \phi(\infty)  =1,\,\,\,  \phi(0)  =0,\,\,\,  A(\infty)   =-1/r,$ 
 where   ${\hat S}$  is an $SU(2)$  subgroup 
 {\small 
$$   S^1= { 1\o 2}   \pmatrix{  0 & 0& 1  \cr  0 & 0 & 0 \cr  1  &0& 0   }; \quad  
S^2= { 1\o  2}   \pmatrix{  0 & 0& -i  \cr  0 & 0  & 0 \cr i &0& 0    }; \quad  
S^3    = { 1\o 2}   \pmatrix{  1  & 0  & 0  \cr  0 & 0 & 0 \cr  0&0& -1   }
$$  }   
or an analogous one in the $(2-3)$  raws and columns. 
 So in this case  there are     two  {\it  degenerate   }   $  SU(3)$ solutions. 
 
The generalization to the case with symmetry breaking 
  {\small $$  SU(N+1)  \to   {SU(N)  \times U(1)  \o {\mathbb Z}_N },
\label{Symbr}$$
$$  \bra \phi  \ket  =  
 \pmatrix{  v     &  0  &  \ldots &    0
  \cr  0   & v    & \ldots   & 0
 \cr  \vdots  &\vdots& \ddots & \vdots
\cr  0    & 0   &  \ldots  & -N v     } =
\pmatrix{  v \cdot {\bf 1}_{N\times N}     &
  \cr    & -N v     },  
\label{phivev}  $$  } 
is straightforward.    
   Consider  a broken $SU(2)$, $S_i$ living in $(1, N+1)$ rows/columns: then
  {\small $$
\phi=\pmatrix{  -\frac{N-1}{2} v     &  0  &  \ldots &    0
 &0
  \cr  0   & v    & 0 & \ldots    &0
\cr 0 & 0 & v & \ldots & 0
 \cr  \vdots  &\vdots& \vdots& \ddots & \vdots
\cr  0    & 0  &0 &  \ldots  & -\frac{N-1}{2} v   }
+(N+1)v  (\vec{S} \cdot \widehat{r}) \phi(r),
$$  
$$ 
\vec{A}(r)=\vec{S} \wedge \widehat{r}  \, \frac{A(r)}{g}    $$  }
   gives a monopole  solution of $SU(N+1)$ equations of motion. By considering various 
   $SU(2)$  subgroups living in  $(i, N+1)$ rows/columns, $i=1,2,\ldots, N$, 
one is led to   $N$ degenerate   solutions.

\section{ Homotopy Groups in Sytems    $G \to H$   }
  
\begin{figure}
\begin{center}
\includegraphics[width=3 in]{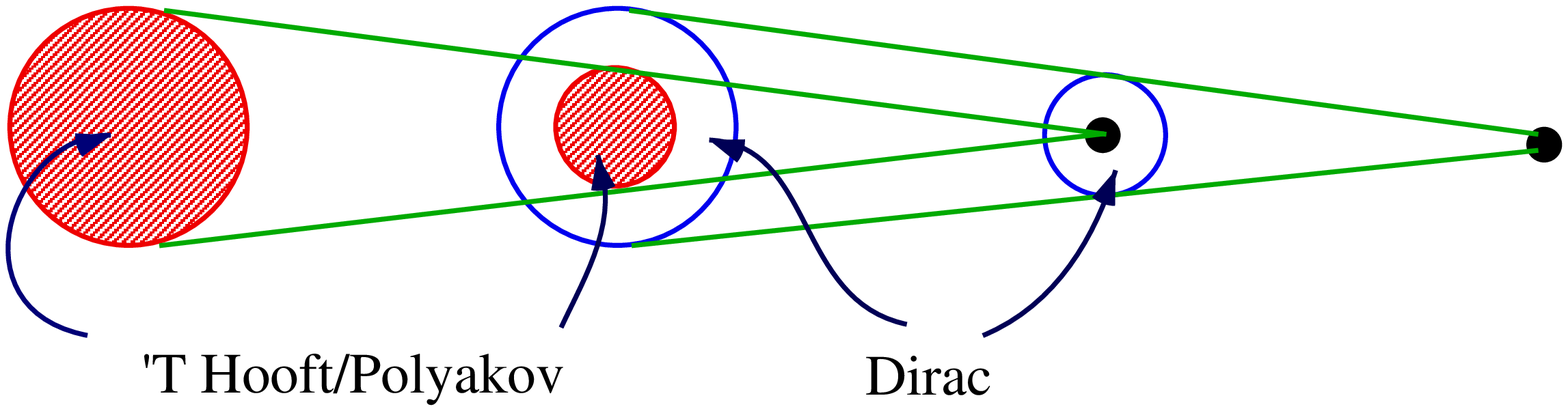}
\caption{ }
\label{homotopy}
\end{center}
\end{figure}

  Let us consider now the relevant  homotopy groups.  
The  short exact sequence 
\[
  \label{eq:1b}
  0 \rightarrow \pi_2 (G/H) \stackrel{f}{\rightarrow} \pi_1 (H) \rightarrow \pi_1 (G) \rightarrow 0.
\]  
tells us that   regular  (BPS)      monopoles   represent   $\pi_2 ( G/H)  \subset   \pi_1(H)   \subset  \pi_1(G) $.
 Alternatively  (Coleman)
one can say that regular  monopoles  correspond to the    kernel of mapping   $\pi_1(H) \to \pi_1(G).$   In general, BPS monopoles belong to  a $k$ th tensor irrep of ${\tilde H}$, $k \in \pi_1(H)$. 
The relation between  't Hooft-Polyakov (regular) monopoles and Dirac (singular) monopoles is illustrated in Figure \ref{homotopy}, which schematically represents the exact sequence above.

\section{ Monopoles are multiplets of ${\tilde H}$}

A crucial  fact for us  is that   {\it  monopoles are multiplet of ${\tilde H}$ }  and not of the original gauge group $H$. 
This is  most clearly seen in the case of  $USp(2N+2 )  \to  USp(2N) \times  U(1)  $   where we find  
  $2 N+1$  degenerate monopoles  (of  $ \tilde{USp}(2N) = SO(2N+1)$ !),  or in the system
   $SO(2n+3) \rightarrow SO(2n+1) \times U(1)$, where the multiplicity of degenerate monopoles is $2N$  (a right
   number for the fundamental representation of      $ {\tilde{ SO}} (2N+1)= {USp}(2N).$)
  
We have recently re-examined the possible irreducible representation (of the dual group  ${\tilde H}$)  to which 
  monopoles belong, in various cases.  The results are shown in Table \ref{tablemono} taken from \cite{ABEKM}.
  
 \begin{table}[h] 
\begin{center}
  {\footnotesize  {
\begin{tabular}{| c | c | c  | c |  c |   }
\hline $G$  & $H$ & ${\tilde  H}$  &   Irrep & { $U(1)$}
\\  \hline  $SU(N+1)$ & ${SU(N) \times U(1)\o  {\mathbb Z}_N} $  &
$SU(N)\times U(1)$  &  $ { N}$    &      $1/N$
\\  \hline  *   ~ $SU(N)$    &  ${SU(r) \times U(1)^{N-r+1}\o  {\mathbb Z}_r}$       &
$ SU(r) \times U(1)^{N-r+1}$   & $ { r}  $     &      $1/r$     
\\  \hline    $USp(2N+2) $ &   $USp(2N) \times
U(1)  $      & $SO(2N+1)\times U(1)$  & $ { {2N+1}}$     & $1 $
\\  \hline  *   ~ $USp(2N+2)$     &  ${SU(r) \times U(1)^{N-r}\o  {\mathbb Z}_r}$       &
$ SU(r) \times U(1)^{N-r+1}$   & $ { r}  $     &      $1/r$    
\\  \hline $SO(2N+3) $ &  $ SO(2N+1) \times U(1)$  
        &  $USp(2N)\times U(1)$ &   $ { {2N}} $  &   $1  $
\\  \hline  $SO(2N+2)$   &   $ SO(2N) \times U(1)$
       &  $SO(2N)\times U(1)$  &  $ { {2N}}$  &    $1$  
\\  \hline  $USp(2N) $ & $  {SU(N) \times   U(1) \o {\mathbb Z}_N}$    
     & $ SU(N)\times U(1)$ &  $ {  N} $   &      $1/N$      
\\  \hline $SO(2N)$   &   ${ SU(N) \times   U(1)\o{\mathbb Z}_N}$      &
  $SU(N)\times U(1)$  & $ { {N(N-1) \o 2} }$     &      $2/N$      
\\  \hline  *   ~ $SO(2N)$    &  ${SU(r) \times U(1)^{N-r+1}\o {\mathbb Z}_r}$       &
$ SU(r) \times U(1)^{N-r+1}$   & $ { r}  $     &      $1/r$     
\\  \hline  $SO(2N+1)$   &    $  {SU(N)\times   U(1) \o  {\mathbb Z}_N }$  &
 $SU(N)\times U(1)$  & $ { {N(N+1) \o  2} }$        &
 $2/N $      
 \\  \hline  * ~ $SO(2N+1)$   &    ${SU(r) \times U(1)^{N-r+1}\o  {\mathbb Z}_r}$      &
 $ SU(r) \times U(1)^{N-r+1}$   & $ { r}  $     &      $1/r$ 
\\  \hline $SU(N+M)$ & ${SU(N)\times SU(M) \times U(1) \o  {\mathbb Z}_k}$ &
 $SU(N)\times SU(M) \times U(1)$ & $({
 N},{\overline M})$ & $1/k$ 
\\  \hline $SO(2N+2M)$ & $SO(2N)\times U(M)$ & $SO(2N)\times U(M)$ &
$({2N}, {M})$ & $1/M$
\\  \hline $SO(2N+2M+1)$ & $SO(2N+1)\times U(M)$ & $USp(2N)\times U(M)$ &
$({2N}, {M})$ & $1/M$
\\  \hline $USp(2N+2M)$ & $USp(2N) \times U(M)$ & $SO(2N+1) \times U(M)$ &
$(2N+1, M)$ & $1/M$
\\ \hline
  \end{tabular}
}} 
\end{center}
\caption{\small  Stable nonabelian magnetic monopoles of minimum
    mass arising from the breaking $G \to H$ and     their
    charges.  The $U(1)$ magnetic charge is given in the unit of the
    minimum Dirac quantum, $ 1/2 \, e_0$, where $e_0$ is the minimum
    electric $U(1)$ charge in the system.  $k$ in the $SU(N+M)$   case    is the least common
    multiple of $N$ and $M$.  The cases with  nonmaximal nonabelian factors (*), $r <N $,    are  qualitatively different from the  
    case with $r=N$  in that  monopoles in the fundamental as well as in the  second-rank tensor representation of $SU(r)$ appear: 
 only the monopoles
in the fundamental representation  of $SU(r)$ are indicated.   Note that  these     do not exist for 
  $r=N$ in the case of   $G=SO(N)$, as most easily  seen from the explicit construction.    }
\label{tablemono}
 \end{table}

\section{    Why Nonabelian Monopoles are Intrinsically Quantum Mechanical  }    

Nonabelian monopoles turn out to be  essentially quantum mechanical. 
In fact, 
finding semiclassical  degenerate   monopoles, as reviewed above,  is  not sufficient for us  to conclude that they form   a multiplet of ${\tilde H}$,  as     $H$ can break itself  dynamically 
at lower energies  and break the degeneracy among the monopoles.  We must ensure that this  does not take place.  Nonabelian  monopoles  are in this sense   never  really semi-classical, even if    $\bra \phi \ket  \gg \Lambda_H:$     ({{\it e.g.,}   Pure $ N=2$,    $SU(3)$  }).

In this connection,  there is a famous ``no go theorem'' which states that there are no ``colored dyons"\cite{CDyons}.
For instance,   in the background of the monopole arising from the breaking  {$SU(3)\o SU(2)\times U(1)$},     no  global  $T^1, T^2,   T^3$ isomorphic to $SU(2)$ can be shown to  exist.  
 These  results have somewhat obscured  the whole issue of nonabelian monopoles for some time. Do they not exist?    Are they actually inconsistent?  
 The way out  of this impasse is actually very  simple:  nonabelian  monopoles  are multiplets of the dual ${\tilde H}$  group,  and the results of \cite{CDyons}  does  not exclude  existence of  sets of monopoles transforming as members of a dual multiplet  (even if at present the explicit form of  such nonlocal transformations are not known; see however below). 

   Nevertheless, the no-go theorem  implies  that the true gauge group of the system  is not 
$$  G_{gauge} \ne   H \otimes   {\tilde H}  
$$  as sometimes suggested, but  $H$  or ${\tilde H}$ or something else, according to which degrees of freedom 
 are effectively present. (See also \cite{BS}).

\section {Phases of Softly Broken ${ N}=2$ Gauge Theories}

Fully quantum mechanical results about the phases  of $SU(n_{c})$,  $USp(2n_{c})$ and $SO(n_{c})$
 theories  with $n_{f}$ hypermultiplets  (quarks), perturbed by the superpotential
 \[
{ W} (\phi, Q, {\tilde Q}) =   \,\mu  \,\Tr  \Phi^2  
 +      m_i    {\tilde Q}_i Q^i, \qquad   m_i \to 0    
\]      
are known \cite{APS,CKM}. (See Table).   

\begin{table}[h]

\begin{center}
  {\small 
\begin{tabular}{cccc}
    Deg.Freed.      &  Eff. Gauge  Group
&   Phase    &   Global Symmetry     \\
\hline
   monopoles   &   $U(1)^{n_c-1} $               &   Confinement
   &      $U(n_f) $            \\ \hline
  monopoles         & $U(1)^{n_c-1} $        &
Confinement       &     $U(n_f-1) \times U(1) $        \\ \hline
 NA monopoles      &    $SU(r)
\times U(1)^{n_c-r}   $  &    Confinement
&          $U(n_f-r) \times U(r) $
\\ \hline
  rel.  nonloc.     &    -    &    Confinement
&          $U({n_f / 2} ) \times U({n_f/2}) $          
\\ \hline
 NA monopoles     &
$ SU({\tilde n}_c) \times  U(1)^{n_c -  {\tilde n}_c } $                &
Free Magnetic
&      $U(n_f) $         \\ \hline
\end{tabular}  }
\caption{ { Phases of $SU(n_c)$ gauge theory with $n_f$ flavors.       
$ {\tilde n}_c
\equiv n_f-n_c$. } }  
\label{tabsun}
\end{center}
\end{table}

\begin{table}[h]

\begin{center}
  {\small 
\begin{tabular}{cccc}
   Deg.Freed.      &  Eff. Gauge Group
&   Phase    &   Global Symmetry     \\
\hline
  rel.  nonloc.       &    -    &
Confinement
&          $ U(n_f)  $
\\ \hline
  dual quarks     &      $USp(2  {\tilde n}_c) \times
U(1)^{n_c -{\tilde n}_c} $               &  Free Magnetic
&      $SO(2n_f) $         \\ \hline
\end{tabular}
  }
\caption { Phases of $USp(2 n_c)$ gauge theory  with $n_f$ flavors  with
$m_i \to 0$.   $ {\tilde n}_c \equiv n_f-n_c-2$. }
\label{tabuspn}
\end{center}
\end{table}

From these results  we learn, in particular,   that
 the spectrum of the ``dual quarks''  in the infrared  theory (charges, multiplicity, flavor) is 
  identical to what is expected from the semiclassical  abelian or nonabelian  monopoles. 
We note in particular  that the $r-$ vacua  ({\it i.e.} vacua with a  low-energy effective $SU(r)$  gauge group)  exist    only   for
$   r  <   {n_f \o 2  }$, namely as long as  
  the   sign-flip of the  beta  function occurs:   
\[     b_0^{(dual)} \propto   -  2 \, r  +   n_f  >  0,   
\label{betadual}   \qquad 
     b_{0} \propto  -  2 \, n_c +    n_f  <     0.     
\label{betafund}   \]
Indeed, analogous $r$  vacua exist semiclassically for all values up to $min (n_{f}, n_{c})$,  but quantum mechanically, only those with $r \le n_{f}/2$  give rise to  vacua with nonabelian gauge symmetry.
Also,
  when the sign flip is not possible   (e.g.  ${N}=2$ YM  or on a generic point of the quantum moduli space)    
dynamical Abelianization  is expected and does take place!    

These observations led us to conclude that  the   ``dual quarks''  belonging to the fundamental representation  of the infrared $SU(r)$  gauge group, actually {\it are }    the Goddard-Nuyts-Olive-Weinberg  monopoles, which have become massless by quantum effects \cite{BK}.

Most importantly, we are  led to the 
  general criterion  for nonabelian monopoles to survive  quantum effects: 
  the system   must  produce, upon symmetry breaking,   a sufficient number of  massless flavors  to protect  $H$ from    becoming too strongly-coupled.  Natural embedding in ${N}=2$ systems  for various cases in Table \ref{tablemono}
has been discussed in Ref.\cite{ABEKM}

A very subtle hint about the nature of the nonabelian monopoles   come from the recent discovery of 
   nonabelian vortices. 

\section {      Vortices        }
  
  Vortices occur in a system where a gauge group $H$ is broken to some discrete group
\[   H   \,\,\,{\stackrel {\bra \phi \ket    \ne 0} {\longrightarrow}}     \,\,\, C \qquad \qquad  \]  
such that  $ \Pi_1( H   /  {C}) $ is not trivial. 
Gauge field behaves far from the vortex axis as 
 \[   A_i  \sim   { i \o  g}    \, U(\varphi) \de_i  U^{\dagger}(\varphi);    
\quad     \phi_A \sim   U \phi_{A}^{(0)}  U^{\dagger},   \qquad  U(\varphi) = \exp{i \sum_j^r 
\beta_j T_j
\varphi} \]
   Quantization condition reads 
  $ \, {\mathbf \alpha }  \cdot {\mathbf \beta  } \subset {\mathbb Z}$ where     
$ \beta_i $  are    weight vectors  of  $ {\tilde H },  $  dual   of $H$   
    Some known  cases are: 
     \begin{itemize}
\item  $H= U(1)$:   in this case    vortices correspond to the well-known Abrikosov-Nielsen-Olesen 
vortices,   representing elements of  $\Pi_1(U(1)) = {\mathbb Z}.$
According to the parameters appearing in the system they yield Type I, Type II or  BPS   superconductors; 

\item   The case  $H= SU(N) /  {\mathbb Z}_N$  yields      $ {\mathbb Z}_N$  vortices.  These are   
non  BPS and are  difficult to analyse   (model dependence),  although there are some interesting work on       
 the tension ratios, the sine formula ($  T_k  \propto     \sin { \pi \, k \o N}$), etc.\cite{DS}
 
\end{itemize}

\section { Nonabelian Vortices }
 
Truely nonabelian   vortices  ({\it i.e., }  with a nonabelian flux)  have recently been  constructed \cite{HT,ABEKY}. 
In the simplest case,  we  consider   the system 
\[    SU(3)    \,\,\,{\stackrel {v_1}{\longrightarrow}}     \,\,\, {SU(2) \times U(1) \o  {\mathbb Z}_2  }    \,\,\,{\stackrel
{v_2}{\longrightarrow}}     \,\,0,
\qquad  v_1 \gg v_2,
\]
 The high-energy  theory      has     monopoles;     
 the low-energy theory     (monopoles heavy)  has  vortices. 
We are here mainly interested in  the low-energy theory  ($ {SU(2) \times U(1) \o  {\mathbb Z}_2  }    \,\,\,{\stackrel
{v_2}{\longrightarrow}}     \,\,0,$). 
We embed the system in a 
    ${N}=2$  model  with  number of flavor,   $4 \le n_f \le  5,$   so as   to
 maintain  the  ``unbroken''  subgroup  $SU(2)$  non asymptotically free. We shall
 take the    
 bare mass $m$   and the adjoint scalar  mass $\mu \, \Phi^2$, 
 so that   $v_{2}=\xi=\sqrt{\mu \,m } \ll   v_{1}=m.$  The scalar VEVs are 
\be
\Phi = -{1\over\sqrt{2}}\left(
\begin{array}{ccc}
  m & 0 & 0 \\
  0 & m & 0 \\
  0 & 0 & -2m
\end{array}  \right),  \,\,  <q^{kA}>=<\bar{\tilde{q}}^{kA}>=\sqrt{\frac{\xi}{2}}\left(
\begin{array}{cc}
  1 & 0  \\
  0 & 1  \\
  \end{array} \right), \label{VEVS}
\ee
 where only nonvanishing color (vertical) and flavor (horizontal) components of squarks  are shown. 
Set $\Phi= \bra \Phi \ket $;    $ q = {\tilde q}^{\dagger}$;   and $ q \to { 1\o 2}  q $, then the action density is
\[
 \left[\frac1{4g^2_2} \left(F^{a}_{\mu\nu}\right)^2 +
\frac1{4g^2_1}\left(F^{8}_{\mu\nu}\right)^2
+ \left|\nabla_{\mu}
q^{A}\right|^2
+ \frac{g^2_2}{8}\left(\bar{q}_A\tau^a q^A\right)^2   +
\frac{g^2_1}{24}\left(\bar{q}_A q^A - 2\xi
 \right)^2
\right].
\label{le}
\]

\section {Nonabelian Bogomolnyi Equations}   

The tension reads 
\[
T \int{d}^2 x   \left [   \sum_{a=1}^3    A_{a}^{2 } + B^{2}  +\frac{1}{2} |C|^{2} +  \frac{\xi}{2  \sqrt{3}}\tilde{F}^{(8)}  \right], \]
\[
A_{a}=   \frac1{2g_2}F^{(a)}_{ij } \pm
     \frac{g_2}{4}
\Big(\bar{q}_A\tau^a q^A  \Big)   \epsilon_{ij},   \qquad  B=  \frac1{2g_1}F^{(8)}_{ij} \pm
     \frac{g_1 }{4\sqrt{3}} 
\left(|q^A|^2-2\xi \right)
\epsilon_{ij },\]
\[   C= \nabla_i \,q^A \pm   i   \epsilon_{ij}
\nabla_j\, q^A, \]
leading to the nonabelian Bogomolnyi Equations, 
$A_{a}=B=C=0.$     The  vortex flux    ($SU(N)$)   is 
\[  {\vec B} =  \nabla     \wedge   \vec{A}, \qquad   
 F_v =   \int_{R^2}  d{\bf S}  \cdot { \Tr  \,  \phi \,  {\bf  B} \o { 1\o \sqrt{2}}  (\Tr  \,  \phi \phi)^{1/2}   }  = 
2 \pi   \cdot \sqrt{2 (N+1) \o N  }, \label{Vorfluxbis} \] 
This matches exactly   the monopole flux Eq.~(\ref{monoflux}) \cite{monovort}. A crucial fact is that there is an unbroken   global  symmetry,  $SU(2)_{C+F}$ (see Eq.~(\ref{VEVS})), broken only by 
the  vortex  configuration (to a $U(1)$).  This implies the existence of  exact zeromodes (moduli)  labelling  
$$  {SU(2) / U(1)}  =   S^2 =  {\bf CP}^1. $$ 
  
  {\bf The vortex  of Generic Orientation (Zero Modes)  }
  can be explicitly constructed  as 
 \bea
&&    q^{kA}=U\left(
\begin{array}{cc}
 e^{  i \varphi}   \phi_1(r) & 0  \\
  0 &  \phi_2(r) \\
  \end{array}\right)U^{-1}=   e^{\frac{i}{2}\varphi  (1+n^a\tau^a)} \,  U\left(
\begin{array}{cc}
  \phi_1(r) & 0  \\
  0 &  \phi_2(r) \\
  \end{array}\right)U^{-1},
\non \\
&&   {\bf A}_{i}(x) = U  [- {\tau^3\o  2} \, \epsilon_{ij}\,\frac{x_j}{r^2}\,
[1-f_3(r)] ]   U^{-1} = -\frac12\,n^a \tau^a\epsilon_{ij}\,\frac{x_j}{r^2}\,
[1-f_3(r)],
\non \\  
&&   A^{8}_{i}(x) = -\sqrt{3}\ \epsilon_{ij}\,\frac{x_j}{r^2}\,
[1-f_8(r)]
 \label  {Vorttr} \non  \eea   
where 
$$   U  \in   SU(2)_{C+F}, \qquad    U \, \tau^3 \, U^{\dagger} =   n^a \, \tau^a,  $$
$$n^a=(\sin \alpha \cos \beta,  \sin \alpha  \sin \beta , \cos \alpha),\qquad   U =e^{ -i \beta\, {\tau_3 /  2} } \,  e^{ -i \alpha  \, {\tau_2 / 2} }. $$  
The tension 
$ T=2\pi\xi  $
is   independent of  $U$. 
  
  \bigskip
  
\noindent  {\bf Remarks: }

 In more general  $SU(N+1)\to  {SU(N)  \times   U(1) \o   {\mathbb Z}_{N } } \to  \emptyset\,\,\,$
systems   with flavors
$2 N +2   >    N_f \ge   2 N,    $  
there appear    vortices    with  $ 2 (N-1)$ - parameter family of   zero modes,
parametrizing     
$$ {SU(N)  \o   SU(N-1) \times    U(1)} \sim  {\mathbb {CP}}^{N-1}: $$
they nicely match the space of (quantum) states of  a particle in the fundamental representation of an $SU(N)$. 
(Actually, for $ N_f >  N$ there are other vortex zero modes (semilocal strings),  not related to the unbroken, exact $SU(N)_{C+F}$  symmetry.  Those are related to  the flat directions.)

Furthermore, vortex dynamics      ($   {SU(2)  \times   U(1) \o   {\mathbb Z}_{2 }}   \to \emptyset  $)
$$   {\bf  n}  \to    {\bf n}(z,t)       $$
can be shown to be equivalent to:
$$
S^{(1+1)}_{\sigma}=\beta \int d^2 x  \,   \frac12 \left(\de \, n^a\right)^2  + {\hbox {\rm fermions}}:   
$$
an $O(3)={\mathbb  CP}^1$ sigma model \cite{NSVZ,VH}!   It has   two vacua;  no spontaneous breaking of $SU(2)_{C+F}$ occurs;  Also, there is  a close 
connection between the 
2D  vortex sigma model  dynamics and the  4D gauge theory  dynamics: they are dual to each other \cite{DHTSY}.

 In  ${N}=2$  theory,  due to  the presence of two independent  scales which we take very different ($\mu \ll   m$),   we can  study   
 monopoles  (HE theory)  and  vortices (LE theory) separately  in the effective theories valid at respective scales.   Physically, of course,  it is perfectly sensible   to consider both together;  the only problem is that it  becomes very  difficult to disentangle the two if the two scales are  of the same order. 

Nevertheless, there remains the fact that 
   monopoles  (of the HE theory)  and  vortices (of the LE theory)     are actually  incompatible   - as  static configurations.
In fact, 
  the  monopoles of HE theory  represent   $\Pi_2({ SU(N+1)  /   SU(N) \times U(1) /  {\mathbb Z}_N  }) $;\,\, but
 in the full theory,  $\Pi_2({ SU(N+1) })=\emptyset$,    so  no topologically stable monopoles exist. 
On the other hand, 
  vortices represent   $\Pi_1(  {SU(N) \times U(1) \o  {\mathbb Z}_N} )$ 
 of LE theory; no vortices exist  in the full theory  since \,\,  $  \Pi_1(SU(N+1)) =\emptyset.$

 What happens is that monopoles of  $G/H$    are confined by  magnetic vortices of $H \to \emptyset$, leading to   monopole-vortex-antimonopole bound states, which are not stable as static configurations. They could however  give rise to rotating and  dynamically stable states.   After all, the mesons in QCD {\it are}  systems of  this kind!  

Note that the restriction on the number of flavors, 
$2 N +2   >    N_f \ge   2 N,    $  is fundamental.  If $N_{f} $ were  less than $2 N$, the subgroup $SU(N)$ would become strongly coupled, and break itself dynamically. Nonabelian  vortices do  not exist quantum mechanically in such a system. 

We are then led  to the following relation between the vortex zeromodes 
and  ${\tilde H}$  transformation of monopoles. Consider a configuration consisting of a monopole (of  $G/H$)
and an infinitely long vortex, which carries away the full monopole flux. At small  distances $r$ from  the monopole center, $r  \sim  O(1/m)$, HE theory is a good approximation  and  the monopole flux looks isotropic; at a much larger distances 
of order of,  $r >  { 1\o \xi},$  one sees the vortex of LE theory.  The energy of the configuration is unchanged if the whole system is rotated by the exact $H_{C+F}$  transformation. 
This is a nonlocal transformation. The end point monopole is apparently transformed by the $H_{C}$ part only,  but,  since  in order to keep the energy of the  whole system unchanged   it is necessary to transform the whole system, it is not a simple  
gauge transformation $H$ of the original theory.  
It is in  this sense that the nonlocal, global  $H_{C+F}$  transformations  can be  interpreted as the dual transformation
${\tilde H}$ acting on the monopole, at  the endpoint of the vortex.  

\begin{center}
    \includegraphics[height=2in,width=3in]{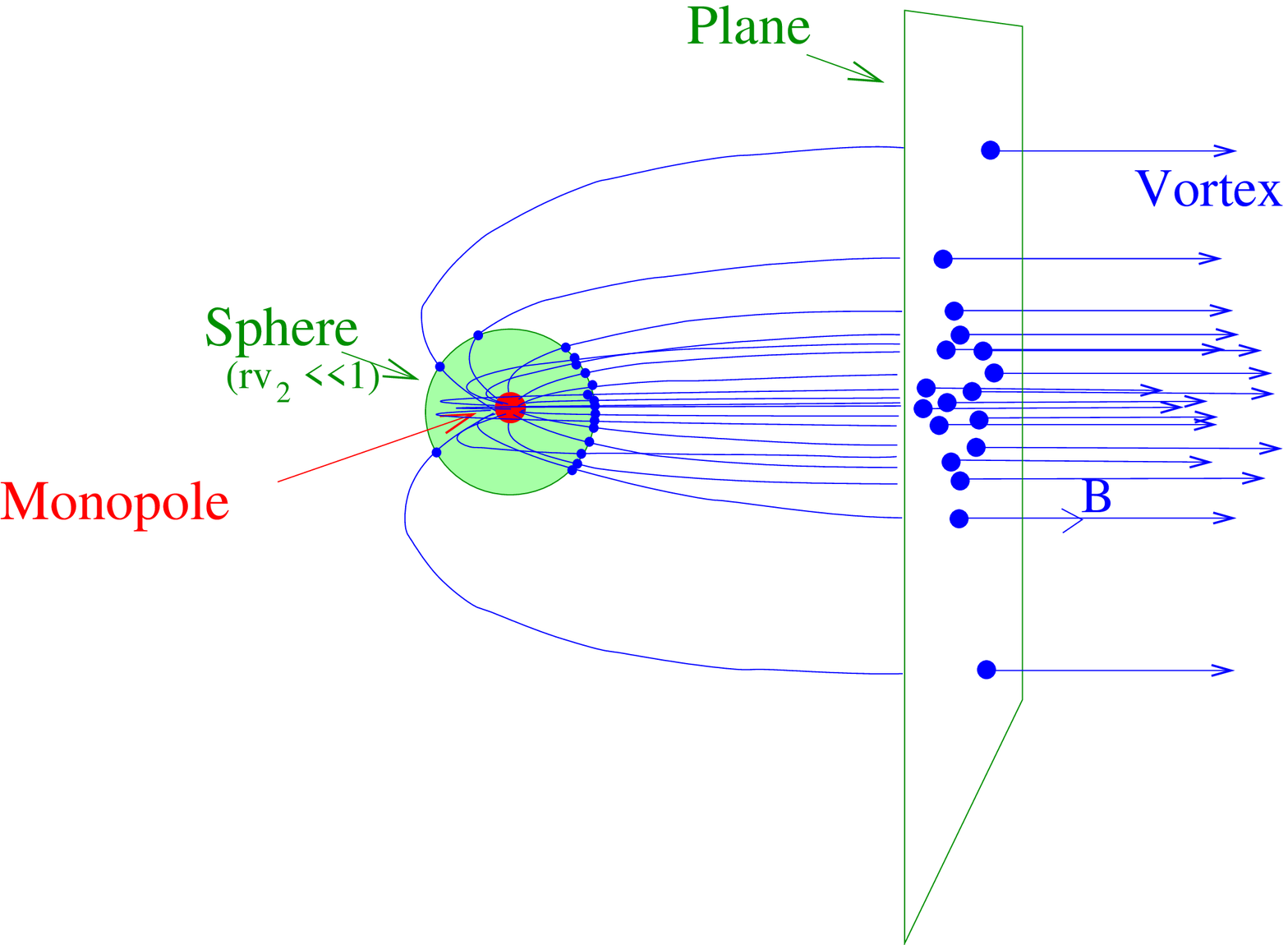}
     \end{center}

\section   { To conclude: where do we stand  ?  }

\begin{itemize}    

\item  Nonabelian monopoles are intrinsically  quantum mechanical;

\item Massless flavors are  important  for (i) keeping $H$ unbroken;
and  (ii)  for providing enough global symmetry giving rise to exact vortex zeromodes:
these can be interpreted as the dual gauge transformation acting on the monopoles at the ends of the vortex; 

\item One has a nice "model" of monopole confinement by vortices. 

\item  Light nonabelian  monopoles appear as IR degrees of freedom
(examples in ${N}=2$  models).   Are there light nonabelian  monopoles in
some  other ${N}=1$ theories? 

\item Do some vacua of  ${N}=2$  theories, especially those based on ``almost superconformal vacua'' \cite{AGK} provide a good model of confinement in QCD?

 \end{itemize}

\section*{Acknowledgment}
I thank Arkady Vainshtein and the organizers of the Workshop for their  kind hospitality  and for providing us with an  occasion for interesting discussions with many participants.


\begin{thebibliography}{100}

\bibitem{tHooft}  G. 't Hooft,  {\bf Nucl. Phys. B138} (1978) 1;  {\it ibid} {\it B153 } (1979) 141, {\it B190} (1981) 455. 

\bibitem{TH}  G. 't Hooft, {\bf  Nucl. Phys.   B190}   (1981) 455;
S. Mandelstam,  {\bf Phys. Lett.  53B }  (1975) 476.

\bibitem{CSW}  F. Cachazo, N. Seiberg and  E. Witten, 
{\bf  JHEP 0302}  (2003)  042,  hep-th/0301006. 

\bibitem{FN}   T.T. Wu and C.N. Yang,  in  ``Properties of Matter Under
Unusual Conditions", Ed. H. Mark and  S. Fernbach, Interscience, New York, 1969,  Y.M Cho, {Phys. Rev.} {\bf D21} 1080 (1980),   L.D. Faddeev and  A.J.
Niemi,  {Phys. Rev. Lett.} {\bf 82}  (1999) 1624, hep-th 9807069. 


\bibitem{DG}  L. Del Debbio, A. Di Giacomo and G. Paffuti,  {Nucl. Phys. B} (Proc. Suppl.) {\bf   42}
(1995) 231; A. Di Giacomo,  hep-lat/0206018.  

 \bibitem{CDSW}  R. Dijkgraaf and    C.Vafa,   hep-th/0208048; F. Cachazo, M. R. Douglas, N. Seiberg and  E. Witten, 
{\bf  JHEP 0212}  (2002)  071,  hep-th/0211170.


\bibitem{WY}  T.T. Wu and C.N. Yang, {\bf Phys. Rev. D12  } (1975) 3845.

\bibitem{TP}    G. 't Hooft,  {\bf Nucl. Phys. B79} (1974) 276;  A.M. Polyakov, {\bf JETP Lett. 20}  (1974) 194.  


\bibitem{Lb}  E. Lubkin, {\bf Ann. Phys. 23}  (1963) 233.

\bibitem {MO} C. Montonen and D. Olive, {\bf Phys. Lett. 72 B} (1977) 117. 

\bibitem{GNO}   P. Goddard, J. Nuyts and D. Olive,   {\bf Nucl. Phys.  B125}  (1977) 1.

\bibitem{BS}  F.A. Bais, {\bf Phys. Rev. D18}  (1978) 1206;   B.J. Schroers and  F.A. Bais, {\bf Nucl. Phys. B512}  (1998) 250,  
hep-th/9708004;    {\bf Nucl. Phys. B535} (1998) 197,  hep-th/9805163. 

\bibitem{EW}   E. J. Weinberg, {\bf Nucl. Phys. B167} (1980) 500;  {\bf Nucl. Phys. B203} (1982) 445. 

\bibitem{SC}   S. Coleman, ``The  Magnetic Monopole  Fifty  Years Later,"
Lectures given at Int. Sch. of Subnuclear Phys., Erice, Italy  (1981). 

\bibitem{Chan} Chan Hong-Mo and Tsou Sheung Tsun,
   {\bf Nucl. Phys. B189}  (1981) 364.

\bibitem{CDyons} A. Abouelsaood, {\bf Nucl. Phys. B226} (1983) 309; P. Nelson and A. Manohar,  {\bf Phys. Rev. Lett. 50}
(1983) 943; A. Balachandran, G. Marmo, M. Mukunda, J. Nilsson, E. Sudarshan and F. Zaccaria,   {\bf Phys. Rev. Lett. 50}
(1983) 1553;  P. Nelson and S. Coleman,  {\bf Nucl. Phys. B227} (1984) 1. 

\bibitem{LWY} K. Lee, E. J. Weinberg and P. Yi,  {\bf Phys. Rev. D 54 } (1996) 6351, hep-th/9605229. . 

\bibitem {CJH}   C. J. Houghton, P. M. Sutcliffe,
{\bf J.Math.Phys.38}  (1997) 5576, hep-th/9708006. 

\bibitem{ABEKM} R. Auzzi, S. Bolognesi,  Jarah Evslin, 
K.  Konishi, H. Murayama,   hep-th/0405070.



\bibitem{APS}
A. Hanany and A. Oz, {\bf  Nucl. Phys. B452} (1995) 283, hep-th/9505075, 
P.~Argyres, M.~Plesser and N.~Seiberg, {\bf  Nucl. Phys. B471} (1996) 159,
hep-th/9603042, 


\bibitem{CKM}
G. Carlino, K. Konishi and H. Murayama,
   {\bf  JHEP   0002}  (2000) 004,     hep-th/0001036;
 {\bf    Nucl. Phys.  B590}  (2000) 37,     hep-th/0005076;
G. Carlino, K. Konishi, Prem Kumar  and H. Murayama,
  {\bf    Nucl. Phys.  B608    }  (2001) 51, hep-th/0104064.


\bibitem{BK}   S. Bolognesi and K. Konishi,  {\bf    Nucl. Phys.  B645 }  (2002) 337, hep-th/0207161.


\bibitem{DS}
M.~Douglas and S.~Shenker, {\bf  Nucl. Phys. B447} (1995) 271-296, hep-th/9503163,
A. Hanany, M. Strassler and A. Zaffaroni,  {\bf Nucl.Phys. B513}   (1998) 87,   hep-th/9707244.
 B.  Lucini and  M. Teper,
{\bf Phys. Rev. D64} (2001) 105019,
 hep-lat/0107007, 
 L. Del Debbio, H. Panagopoulos, P. Rossi and  E. Vicari,  {\bf Phys. Rev. D65} (2002) 021501, hep-th/0106185;
{\bf JHEP 0201} (2002) 009,  hep-th/0111090,  
 C. P. Herzog and I. R. Klebanov,  {\bf Phys. Lett. B526} (2002) 388, 
hep-th/0111078,  R. Auzzi and K. Konishi,  {\bf New J. Phys. 4}  (2002) 59 
 hep-th/0205172. 


\bibitem{HT} A, Hanany and D. Tong,  {\bf JHEP 0307} (2003)   037,   hep-th/0306150.
 
\bibitem{ABEKY} R. Auzzi, S. Bolognesi,  Jarah Evslin, 
K.  Konishi and  A.  Yung, {\bf Nucl. Phys. B} to appear,    hep-th/0307287. 

\bibitem{NSVZ}  V. Novikov, M. Shifman, A. Vainshtein and  V. Zakharov,    {\bf  Phys. Rep C}  116   (1984)  103.

\bibitem{VH}
K. Hori and C. Vafa, hep-th/0002222.


\bibitem{monovort} R.~Auzzi, S.~Bolognesi, J.~Evslin and K.~Konishi, {\bf Nucl. Phys. B686} (2004) 119,
 hep-th/0312233.

\bibitem{DHTSY}  A. Hanany  and D. Tong,  {\bf JHEP 0404} (2004)  066,    hep-th/0403158;
M. Shifman and A. Yung,   hep-th/0403149.


 
 

\bibitem{AGK}  R. Auzzi, R. Grena and K. Konishi,   {\bf Nucl. Phys.    B653 } (2003) 204, hep-th/0211282,
   R. Auzzi and  R. Grena, hep-th/0402213.


\end{thebibliography}
\end{document}